%% file: main.tex
\RequirePackage{coverpage}

\tdtitle{Novel Electromagnetism-Based Radar Propagation Model for 5G and Beyond}
\tdauthor{François De Saint Moulin\IEEEauthorrefmark{1}, Christophe Craeye\IEEEauthorrefmark{2}, Luc Vandendorpe\IEEEauthorrefmark{3}, Claude Oestges\IEEEauthorrefmark{4}}
\source{ICTEAM, UCLouvain - Louvain-la-Neuve, Belgium} 
\tdnumber{045}
\address{Place du Levant 3/L5.03.02, 1348 Louvain-la-Neuve}
\phone{+32 10 47 31 14}
\fax{/}
\email{\{\IEEEauthorrefmark{1}francois.desaintmoulin,
\IEEEauthorrefmark{2}christophe.craeye,
\IEEEauthorrefmark{3}luc.vandendorpe,
\IEEEauthorrefmark{4}claude.oestges\}@uclouvain.be}

\documentclass[conference]{IEEEtran}

\IEEEoverridecommandlockouts

\usepackage[caption=false,font=normalsize,labelfont=sf,textfont=sf]{subfig}
\usepackage{textcomp}
\usepackage{stfloats}
\usepackage{verbatim}
\usepackage{cite}
\usepackage{amsmath,amssymb,amsfonts}
\usepackage{algorithmic}
\usepackage{graphicx}
\usepackage{float}
\usepackage{textcomp}
\usepackage{xcolor}
\usepackage{bbm}
\usepackage{array}
\usepackage{rotating}
\usepackage{gensymb}
\usepackage{comment}
\usepackage[acronym]{glossaries}
\usepackage{multirow}
\usepackage{lscape}
\usepackage{makecell}
\usepackage[inline]{enumitem}

\usepackage{adjustbox}
\usepackage{tikz}
\usetikzlibrary{shapes,arrows}
\usetikzlibrary{positioning}
\usetikzlibrary{decorations.text}

\input{commands}

\def\BibTeX{{\rm B\kern-.05em{\sc i\kern-.025em b}\kern-.08em
    T\kern-.1667em\lower.7ex\hbox{E}\kern-.125emX}}

\begin{document}

\title{Novel Electromagnetism-Based Radar Propagation Model for 5G and Beyond}

\author{\IEEEauthorblockN{François De Saint Moulin\IEEEauthorrefmark{1}, Christophe Craeye\IEEEauthorrefmark{2}, Luc Vandendorpe\IEEEauthorrefmark{3}, Claude Oestges\IEEEauthorrefmark{4}\thanks{François De Saint Moulin is a Research Fellow of the Fonds de la Recherche Scientifique - FNRS.}}\\
\IEEEauthorblockA{ICTEAM, UCLouvain - Louvain-la-Neuve, Belgium\\}
\{\IEEEauthorrefmark{1}francois.desaintmoulin,
\IEEEauthorrefmark{2}christophe.craeye,
\IEEEauthorrefmark{3}luc.vandendorpe,
\IEEEauthorrefmark{4}claude.oestges\}@uclouvain.be\vspace*{-0.2cm}}


\maketitle

\thispagestyle{plain}
\pagestyle{plain}


\begin{abstract}
In order to evaluate the performance of radar and communication systems in future wireless networks, accurate propagation models are needed to predict efficiently the received powers at each node, and draw correct conclusions. In this paper, we present new radar propagation models based on the electromagnetism theory. The target is modelled as a flat or curved square plate to compute the scattered field and derive accurate radar cross section modellings. With a flat square plate, the model makes the link between the radar equation and the geometrical optics propagation model used in ray-tracing applications. It is then applied to popular automotive scenarios within the stochastic geometry framework to observe the impact of such modelling.
\end{abstract}



\section{Introduction}
\input{introduction}

\section{EM-based Radar Propagation in Free Space}
\label{sec:EM}
\input{development}

\section{Application - Automotive Scenario}
\label{sec:application}
\input{numerical_analysis}

\section{Conclusion}
\input{conclusion}


\bibliographystyle{ieeetr}
\bibliography{biblio}

\end{document}

%% file: commands.tex
\renewcommand{\exp}[1]{e^{#1}}

\newcommand{\C}{\text{C}}
\newcommand{\R}{\text{R}}
\newcommand{\F}{\text{F}}

\renewcommand{\d}{\text{d}}

%% file: introduction.tex
Radar and communication systems both exploit the ElectroMagnetic (EM) spectrum with specific objectives. On the one hand, radar systems are deployed in order to gather information about the environment by transmitting a known waveform. By receiving and processing the waveform echoes created by multiple targets, the radar system is able to estimate multiple parameters such as the position, speed and angle of the targets. On the other hand, communication systems are deployed in order to transmit information to other systems in an unknown environment, which is usually also estimated in order to compensate for the generated impairments. \smallskip 

Usually, radar waveforms are transmitted at high carrier frequencies, namely 24-26GHz for automotive applications, while communication waveforms are transmitted at lower carrier frequencies, namely 410MHz-5.9GHz for the LTE frequency bands. In recent years, higher frequency bands have been considered for both the radar and communication functions. For example, the 76-81GHz has become the new standard for recent automotive radar systems. Similarly, the 5G-NR standard extends previous standards to cover new spectrum offering up to 7.125GHz in a first instance, and then other frequency bands at 24.25GHz and 71GHz. In this context, radar and communication systems are converging towards the same frequency bands, leading to an increased interest in joint radar-communication technologies. \smallskip

Nowadays, in order to design radar and communication systems, new radar and communication propagation models are required in view of the new carrier frequencies and targeted applications. These models are then used to estimate the received radar echoes and communication signals powers, which directly impacts the performance of both systems. Their accuracy is therefore crucial in order to draw reliable conclusions about the performance of the systems, for instance by means of stochastic geometry which is a popular framework to obtain closed-form expressions of the average performance metrics in large-scale wireless networks \cite{Hmamouche2021,baccelliV1,baccelliV2}.

\subsection{State of the Art - Propagation Models}
\label{sec:SOTA_prop}
Radar and communication propagation models are usually based on EM theory. From the computed electric and magnetic fields, the received power can be computed. For communication applications, the Friis equation \cite{friis} gives the received power $P_\C$ between two nodes separated by a distance $R$ as 
\begin{equation}
P_\C(R) = \text{EIRP} \: G_r \frac{c^2}{4\pi f_c^2} \: \beta^{-1} \: R^{-\alpha}, \label{eq:Friis}
\end{equation}
where $\text{EIRP}$ is the Effective Isotropic Radiated Power (EIRP), $G_r$ is the receive beamforming gain, $f_c$ is the carrier frequency, $\beta$ is the fixed intercept and $\alpha$ is the path-loss exponent of the path-loss model. For free-space propagation, $\beta = 4\pi$ and $\alpha = 2$. For radar applications, the received power $P_\R$ is usually computed following the radar equation \cite{radar_fund}: 
\begin{equation}
    P_{\R,\text{RCS}}(R) = \text{EIRP} \: G_r \frac{c^2}{4\pi f_c^2} \: \beta^{-2} R^{-2\alpha} \: \sigma, \label{eq:PR_RCS}
\end{equation}
with $\sigma$ the Radar Cross-Section (RCS) of the target. However, this model is only valid in the far field, i.e. when $r >> R_\F = 2 D^2 f_c/c$ with $D$ the dimension of the scatterer. In many applications, as automotive scenarios, this assumption is not always fulfilled. Additionally, the modelling of the RCS is difficult and often inaccurate. This can lead to significantly wrong conclusions about the achieved performance. Another approach is based on the geometrical optics theory, which is used in ray-tracing applications \cite{ray_tracing}. In this case, the received power is given by 
\begin{equation}
    P_{\R,\text{RT}}(R) = \text{EIRP} \: G_r \frac{c^2}{4\pi f_c^2} \: |\rho|^2 \: \beta^{-1} (2R)^{-\alpha}, \label{eq:PR_RT}
\end{equation}
where $\rho$ is the Fresnel reflection coefficient. For a Perfect Electrical Conductor (PEC), it is equal to 1. It is similar to \eqref{eq:Friis} in which the receiver would be located at a distance $2R$. However, owing to the geometrical optics approximations, the targets' property are not taken into account.

\subsection{Related Works in Stochastic Geometry}
Within the stochastic geometry framework, for radar applications, \cite{08464057} analyses the cumulative distribution function of the interference with multiple pulsed-radar devices, \cite{08457255} evaluates the detection probability when multiple obstacles are distributed around the radar systems, and \cite{09580712,09764287} study the radar performance achieved in a cluttered environment. It is also used to analyse the sensing performance in automotive scenarios. As a non exhaustive list, \cite{09119440} evaluates the performance achieved for different radar cross section models. Multiple lanes are considered in \cite{08967012} and \cite{Fang2020}, the former equipped with front- and side-mounted radars with directional antenna patterns, and the latter taking into account the interference from reflections on vehicles in the neighbouring lanes. Instead of using poisson point processes to model the nodes' positions, \cite{07819520} uses a one-dimensional lattice, and \cite{09580432,09266343} uses Matérn hard-core processes, respectively in one or two dimensions. Fine-grained analysis is applied in \cite{09685164,09745306} to evaluate the meta distribution of the Signal to Interference plus Noise Ratio (SINR), enabling to analyse the reliability of the detection at each individual vehicle. Recently, joint radar and communication applications have also been analysed. For automotive applications \cite{09855937} and \cite{09606367} evaluate the cooperative detection range of the system, respectively with spectrum allocation between both functions or with a joint system. \smallskip 

In all these works, \eqref{eq:PR_RCS} is extensively used to model the received radar echo power, whatever the target, the radar-to-target distance or the considered carrier frequency. However, this might lead to misleading conclusions. Therefore, new radar propagation models and RCS modellings are needed to improve the accuracy of the performance analysis.

\subsection{Contributions}
The contributions of this paper are summarised as follows:
\begin{itemize}
\item New radar propagation models and RCS modellings are developed from the EM theory with a finite flat or curved square plate. These models are valid in the near and far field. For the flat square plate, this model unifies the radar equation and the geometrical optics propagation model used in ray-tracing applications. The proposed approach can also be generalised for other target geometries.
\item A simple closed-form approximation of the RCS is proposed to simplify the model.  This approximation is better when the carrier frequency is large, and performs well for automotive radar applications at 76,5GHz. 
\item The new models are applied to an automotive scenario initially studied within the stochastic geometry framework to evaluate and compare the achieved performance in the different case. It has shown that the model should be chosen accurately w.r.t. the radar target to well evaluate the performance of the radar system, since huge difference are observed depending on the model choice and parametrisation.
\end{itemize}

\subsection{Structure of the Paper}
First, the EM theory is used in Section \ref{sec:EM} to compute the scattered fields for targets modelled as flat and curved square PEC plates, and derive the corresponding RCSs. Then, new radar propagation models are proposed in Section \ref{sec:radar_prop_models}, and links are highlighted with existing models in the literature. Finally, the new models are applied to an automotive scenario in Section \ref{sec:application} to observe the impact of the propagation model on the evaluation of the achieved radar performance.

%% file: development.tex
In this section, the EM theory is used in order to model the scattered electric field for a square PEC plate geometry, first plane and then curved. We assume that the transmitter uses a linear antenna, such that the transmitted electric and magnetic fields at a distance $R$ of the antenna can be written as 
\begin{align}
    \vec{E}_i &= -j k \eta L I \: \frac{\exp{-j k R}}{4\pi R} \: F(\theta) \: \hat{e}_\theta = E_i(R,\theta) \: \exp{-j k R} \: \hat{e}_\theta, \label{eq:EF_TX} \\
    \vec{H}_i &= \frac{1}{\eta} \: \hat{e}_u \times \vec{E}_i = \frac{E_i(R,\theta)}{\eta} \: \exp{-j k R} \: \hat{e}_\phi. \label{eq:MF_TX}
\end{align}
In this equation, $k = 2\pi/\lambda$ is the wave number with $\lambda$ the wavelength, $\eta$ is the free-space impedance, $L$ is the antenna length, $I$ is the antenna current, $F(\theta)$ is the normalised antenna radiation pattern, with $\theta$ the elevation angle, equal to 0\degree\: when the direction of propagation is parallel to the antenna, $\hat{e}_\theta$ is the elevation unit vector in a polar coordinate system, and $\hat{e}_u$ is the direction of propagation. These equations are valid when $R >> 2L^2/\lambda$, which is rewritten as $R >> \lambda$ when the antenna length is designed proportional to the wavelength. Owing to the considered frequencies for actual radar systems, this assumption is always valid in the considered scenarios.

\subsection{Stationary Phase Approximation (SPA)}
The SPA is useful to compute the backscattered EM fields for different geometries in the near field \cite{stationary_phase}. Let us consider an integral in the following form:
\begin{equation}
I = \int_\Omega g(x) \: \exp{j\phi(x)} \: \d x,
\end{equation}
where $g$ is a function to integrate and $\phi$ a given phase function. The SPA consists in approximating the phase of $\phi$ by a Taylor expansion around a stationary phase $x_0$, defined such that $\phi'(x_0) = 0$:
\begin{equation}
\phi(x) = \phi(x_0) + \frac{1}{2} \phi''(x_0) (x-x_0)^2 + \mathcal{O}(x^3).
\end{equation}
If the value of $\phi''(x_0)$ is sufficiently large with regards to the boundaries of $\Omega$, the second-order term of the phase will oscillate rapidly, leading to successive cancellation. This also requires the function $F$ to vary slowly, especially in the vicinity of $x_0$. In that case, the integration boundaries can be extended to infinity, and the integral is approximated as 
\begin{align}
I &\approx g(x_0) \: \exp{j\phi(x_0)} \: \int_{-\infty}^{\infty} \exp{j\frac{\phi''(x_0)}{2} (x-x_0)^2} \: \d x \\
&= g(x_0) \: \exp{j\phi(x_0)} \: \sqrt{\frac{j2\pi}{\phi''(x_0)}}.
\end{align}

\subsection{Flat Square Plate Geometry}
\label{sec:finite_square_plate}
\begin{figure}
    \centering
    \includegraphics[width=0.9\linewidth]{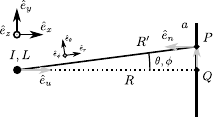}
    \caption{Flat square plate geometry of side length $a$.}
    \label{fig:square_plate}
\end{figure}
Let us consider a flat square plate of side length $a$, assumed to be a perfect electrical conductor. The scenario is illustrated in Figure \ref{fig:square_plate}. The electrical current density generated at the surface of the plate can be computed at every point of the plate based on \eqref{eq:MF_TX}:
\begin{equation}
\vec{J} = 2\:\hat{e}_n \times \vec{H}_i = J(R',\theta,\phi) \: \exp{-jkR'} \: \hat{e}_y,
\end{equation}
with $J(R',\theta,\phi) = \frac{2}{\eta} E(R',\theta)\:\cos\phi$. The backscattered electric field is finally computed from the electrical current density as 
\begin{align}
\nonumber \vec{E}_s &= -jk\eta \int_\Omega \vec{J}_{\perp} \: \frac{\exp{-jkR'}}{4\pi R'} \: \d \Omega \\
&= -jk\eta \int_\Omega \frac{J(R',\theta,\phi)}{4\pi R'} \: \exp{-j2kR'} \: \d \Omega \: \hat{e}_y, \label{eq:scattered_EF}
\end{align}
with $\vec{J}_\perp = \vec{J} - (\vec{J}\cdot\hat{e}_n)\:\hat{e}_n = \vec{J}$ the perpendicular component of the current density, and $\Omega$ the surface of the finite square plate.

\subsubsection{Resolution with the SPA (near field)} 
Let us assume a coordinate system located at the center of the plate, such that $\Omega \triangleq \{(0,y,z) \:|\: y,z \in [-\frac{a}{2},-\frac{a}{2}]\}$. The phase of the integrand is developed as 
\begin{equation}
\phi = -2 k R' = -2 k \sqrt{R^2 + y^2 + z^2}.
\end{equation}
The stationary phase point is also located at the center of the plate. The phase is therefore approximated as 
\begin{equation}
\phi \approx -2 k R \left(1 + \frac{y^2 + z^2}{2R^2}\right).
\end{equation}
In order to apply the stationary phase approximation, the first Fresnel zones must be smaller w.r.t. the plate dimension:
\begin{equation}
    \frac{k}{R} \frac{a^2}{4} >> (2n + 1)\frac{\pi}{2} \Leftrightarrow a >> \sqrt{m \lambda R}, \label{eq:SPA_cond_no_curvature}
\end{equation}
with $m = 2n+1$ and $n \in \mathbb{N}$. Note that this condition implies that $R << 2a^2/\lambda$, meaning that the reflection occurs in the near field. Under that assumption, using the SPA, \eqref{eq:scattered_EF} is approximated as 
\begin{align}
\nonumber \vec{E}_s &\approx -jk\eta \: \frac{J(R,0,0)}{4\pi R} \: \exp{-j2kR} \: \left(\int_{-\infty}^{\infty} \exp{-jk \frac{y^2}{R}} \: \d y\right)^2 \: \hat{e}_y\\
\nonumber  &= -j 2k \: \frac{E(R,0)}{4\pi R}\: \exp{-j2kR} \: \frac{-j\pi R}{k} \: \hat{e}_y \\
&= jk\eta LI \: \frac{\exp{-jk(2R)}}{4\pi(2R)} \: \hat{e}_y. \label{eq:SEF_RT}
\end{align}
Equation \eqref{eq:SEF_RT} is equivalent to the electric field computed at a distance $2R$ of the antenna. This is the result obtained with the geometrical optics propagation model used in ray-tracing applications, which leads to \eqref{eq:PR_RT}. 

\subsubsection{Resolution in the far field}
\label{sec:plane_plate_taylor}
In the far field, $R >> 2a^2/\lambda >> a/2$, meaning that a Taylor approximation of $R'$ can be made, as with the SPA. Additionally, since the plate dimension is small compared to the distance, one can assume that $R' \approx R$ and that $\theta \approx 0\degree$, $\phi \approx 0\degree$ in the amplitude factors only, leading to the following integral:
\begin{equation}
    \vec{E}_s \approx -jk\eta \: \frac{J(R,0,0)}{4\pi R} \: \exp{-j2kR} \: \left(\int_{-\frac{a}{2}}^{\frac{a}{2}} \exp{-jk \frac{y^2}{R}} \: \d y\right)^2 \: \hat{e}_y.
\end{equation}
Since the integration domain cannot be extended to infinity in that case, the integral can be solved using a (normalised) Fresnel integral as done in \cite{9066895}, defined in the complex case as
\begin{equation}
F(x) \triangleq \int_{0}^x \exp{j\pi\frac{t^2}{2}} \:\d t.
\end{equation}
Using the property $2F(x) = F(x) - F(-x)$, the following equality holds:
\begin{equation}
\int_{-A}^{A} \exp{-jBt^2} \:\d t = \sqrt{\frac{2\pi}{B}} \: F^*\left(\sqrt{\frac{2B}{\pi}}A\right),
\end{equation}
leading to the following expression for the backscattered electric field:
\begin{align}
    \nonumber \vec{E}_s &\approx -2jk\eta \: \frac{J(R,0,0)}{4\pi R} \: \exp{-j2kR} \: \frac{\pi R}{k} \left(F^*\left(\sqrt{\frac{2k}{\pi R}} \frac{a}{2}\right)\right)^2 \: \hat{e}_y\\
    &= - k \eta LI \: \frac{\exp{-jk (2R)}}{4\pi (2R)} \: \Gamma\left(\frac{R}{R_\text{F}}\right)\: \hat{e}_y, \label{eq:SEF_general}
\end{align}
where $R_\text{F} = 2a^2/\lambda$ is the Fraunhoffer distance of the scatterer, and $\Gamma$ is defined as 
\begin{equation}
\Gamma(x) \triangleq 2\left(F^*\left(\sqrt{\frac{1}{2x}}\right)\right)^2.
\end{equation}

\subsubsection{Putting everything together}

\begin{figure}
    \centering
    \includegraphics[width=0.9\linewidth]{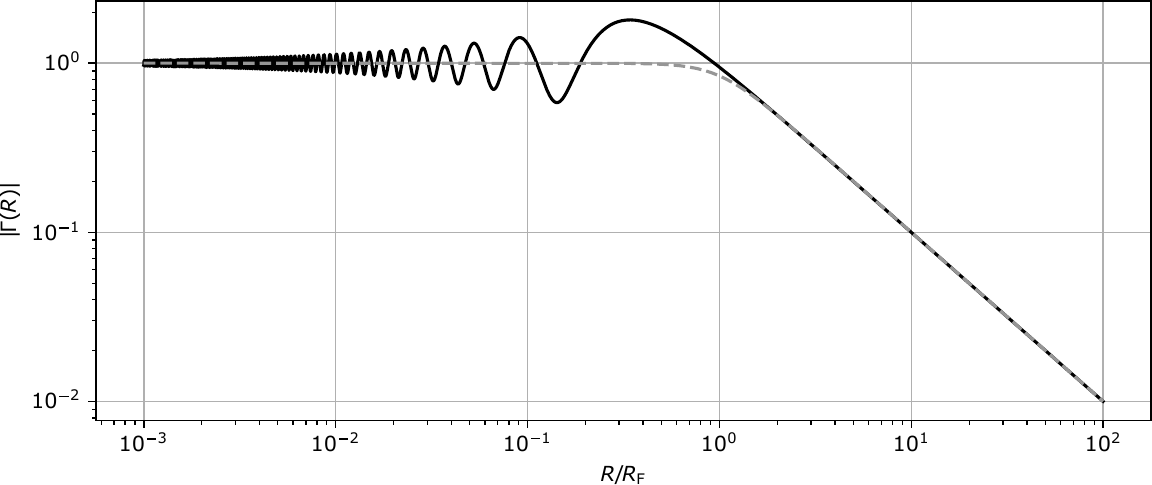}
    \vspace*{0.2cm}
    
    \includegraphics[width=0.9\linewidth]{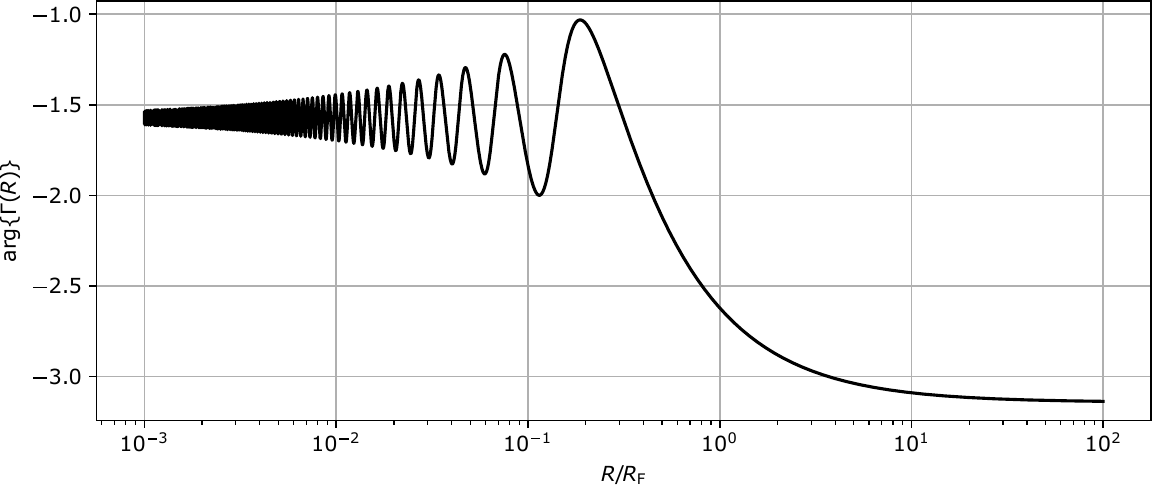}
    \caption{Modulus and angle of the function $\Gamma$ (continuous line), and an approximation for the modulus (dashed line, \eqref{eq:gamma_approx} with $n=4$).}
    \label{fig:gamma}
\end{figure}

Figure \ref{fig:gamma} illustrates the modulus and the angle of the Gamma function. The following approximation for the modulus is also shown:
\begin{equation}
    \left|\Gamma(x)\right| \approx \left(\frac{1}{1+x^n}\right)^{\frac{1}{n}}, \label{eq:gamma_approx}
\end{equation}
with $n \in \mathbb{N}$ and $n \geq 1$. This approximation enables to simplify the model and filter the oscillatory behaviour of the Gamma function. In the near field ($R << R_\text{F}$), $\Gamma(x) \approx -j$, and \eqref{eq:SEF_RT} is obtained again. At the opposite, in the far field ($R >> R_\text{F}$), $\Gamma(x) \approx -R_\text{F}/R$, giving 
\begin{equation}
    \vec{E}_s \approx k \eta LI \: \frac{\exp{-j2kR}}{4\pi R^2} \: \frac{R_\text{F}}{2} \: \hat{e}_y. \label{eq:SEF_RCS}
\end{equation}

One can also compute the scattered power density as 
\begin{align}
S_s &= \frac{|\vec{E}_s|^2}{\eta} = \eta \left(\frac{k L I}{4\pi (2R)}\right)^2 \: \left|\Gamma\left(\frac{R}{R_\text{F}}\right)\right|^2 \\
&= \frac{\text{EIRP}}{4\pi (2R)^2} \: \left|\Gamma\left(\frac{R}{R_\text{F}}\right)\right|^2,
\end{align}
where the EIRP is equal to $\text{EIRP} = \eta k^2 L^2 I^2 / 4\pi$. The incident power density at the center of the plate being computed as 
\begin{equation}
S_i = \frac{E_i^2(R,0)}{\eta} = \frac{\text{EIRP}}{4\pi R^2},
\end{equation}
the RCS is therefore computed as follows: 
\begin{equation}
\sigma = 4 \pi R^2 \: \frac{S_s}{S_i} = \pi R^2 \left|\Gamma\left(\frac{R}{R_\text{F}}\right)\right|^2 \xrightarrow{R >> R_\text{F}} \pi R_\text{F}^2. \label{eq:RCS_value_FS}
\end{equation}
In the near field, $|\Gamma(R/R_\text{F})|^2 \approx 1$, and the RCS is a function of the distance: $\sigma = \pi R^2$. In the far field, $|\Gamma(R/R_\text{F})|^2 \approx R_\text{F}^2/R^2$, and the RCS is a constant: $\sigma = \pi R_\text{F}^2$. \smallskip

\subsection{Extension with curvature}
\begin{figure}
\centering
\includegraphics[width=0.9\linewidth]{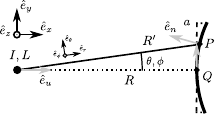}
\caption{Curved square plate geometry of side length $a$.}
\label{fig:curvated_square_plate}
\end{figure}

Let us now consider a curved square plate of side length $a$. In that case, the distance $R'$ is difficult to express as a function of $y$ and $z$. Therefore, we assume that this curved plate can be well approximated by a paraboloid with curvature radius $C_y$ and $C_z$ in the $y$ and $z$ directions, such that
\begin{equation}
R' \approx R + \frac{y^2}{2}\left(\frac{1}{R} + \frac{1}{C_y}\right) + \frac{z^2}{2}\left(\frac{1}{R} + \frac{1}{C_z}\right).
\end{equation}
This scenario is illustrated in Figure \ref{fig:curvated_square_plate}. \smallskip

Let us consider in a first time the case where $C_y = C_z = C$ in order to simplify the discussions. The SPA can be applied if
\begin{equation}
    k \left(\frac{1}{R}+\frac{1}{C}\right) \frac{a^2}{4} >> m \frac{\pi}{2} \Leftrightarrow a >> \sqrt{\frac{m\lambda}{\frac{1}{R}+\frac{1}{C}}}.
\end{equation}
When the curvature radius is large compared to the considered distances, this condition reduces to \eqref{eq:SPA_cond_no_curvature}. In the other case, the condition depends only on the curvature radius:
\begin{equation}
    C << \frac{2 a^2} \lambda \Rightarrow a >> \sqrt{m \lambda C}.
\end{equation} 
Therefore, the results obtained using the SPA can always be applied, except when $R_\F << C << \infty$ and $R >> R_\F$. However, when the SPA cannot be used, one can again rely on the results obtained through Taylor approximation of the distance as done in Section \ref{sec:finite_square_plate} as a good approximation. Following these  developments, the scattered electric field is computed as 
\begin{align}
\vec{E}_s &\approx -jk\eta \: \frac{J(R,0,0)}{4\pi R} \: \exp{-j2kR} \: \frac{\pi \tilde{R}}{k} \: \Gamma\left(\frac{\tilde{R}}{R_\text{F}}\right) \: \hat{e}_y\\
&= - k \eta L I \frac{\exp{-jk(2R)}}{4\pi (2R)} \: \frac{\tilde{R}}{R} \: \Gamma\left(\frac{\tilde{R}}{R_\text{F}}\right)\: \hat{e}_y,
\end{align}
with $\tilde{R} = (1/R + 1/C)^{-1}$. The RCS is then computed as 
\begin{equation}
    \sigma = 4\pi R^2 \: \frac{|\vec{E}_s|^2}{|\vec{E}_i|^2} = \pi \tilde{R}^2 \left|\Gamma\left(\frac{\tilde{R}}{R_\text{F}}\right)\right|^2.
\end{equation}
This shows that, even for distances $R$ smaller than the Fraunhoffer distance $R_\text{F}$, when the curvature is sufficiently high such that $R >> C$, the RCS is constant. In summary, 
\begin{itemize}
\item When $C >> R_\text{F}$ and $C \approx R_\text{F}$, the RCS behaviour is similar to the one obtained with a flat square plate.
\item When $C << R_\text{F}$, either $R << C$ and $\sigma = \pi R^2$, or $R >> C$ and $\sigma \approx \pi C^2 \: | \Gamma(C/R_\text{F})|^2$, even if $R$ is smaller than the Fraunhoffer distance. 
\end{itemize}

Let us now generalise and consider the case where $C_y \neq C_z$. We assume that the SPA can be applied, meaning that
\begin{equation}
    a >> \sqrt{\frac{m\lambda}{\frac{1}{R}+\frac{1}{C_y}}}\quad\text{and}\quad a >> \sqrt{\frac{m\lambda}{\frac{1}{R}+\frac{1}{C_z}}}.
\end{equation}
Following again the developments of Section \ref{sec:finite_square_plate}, the scattered field is computed as 
\begin{equation}
    \vec{E}_s \approx - k \eta L I \frac{\exp{-jk(2R)}}{4\pi (2R)} \: \sqrt{\frac{\tilde{R}_y\tilde{R}_z}{R^2} \: \Gamma\left(\frac{\tilde{R}_y}{R_\text{F}}\right)\Gamma\left(\frac{\tilde{R}_z}{R_\text{F}}\right)}\: \hat{e}_y, \label{eq:SEF_new}
\end{equation}
with $\tilde{R}_y = (1/R + 1/C_y)^{-1}$ and $\tilde{R}_z = (1/R + 1/C_z)^{-1}$, and the RCS is computed as 
\begin{equation}
\sigma = \pi \: \tilde{R}_y\tilde{R}_z \: \left|\Gamma\left(\frac{\tilde{R}_y}{R_\text{F}}\right)\Gamma\left(\frac{\tilde{R}_z}{R_\text{F}}\right)\right|. \label{eq:RCS_curved_plate}
\end{equation}
Finally, in order to simplify the model, the approximation of the $\Gamma$ function given in \eqref{eq:gamma_approx} enables to rewrite \eqref{eq:RCS_curved_plate} as
\begin{equation}
    \sigma \approx \pi R_\F^2  \: \left[\bigg(1 + \bigg(\frac{R_\F}{\tilde{R}_y}\bigg)^n\bigg)\bigg(1+\bigg(\frac{R_\F}{\tilde{R}_z}\bigg)^n\bigg)\right]^{-\frac{1}{n}}. \label{eq:RCS_curved_plate_approx}
\end{equation}

\begin{figure}
\centering
\includegraphics[width=0.9\linewidth]{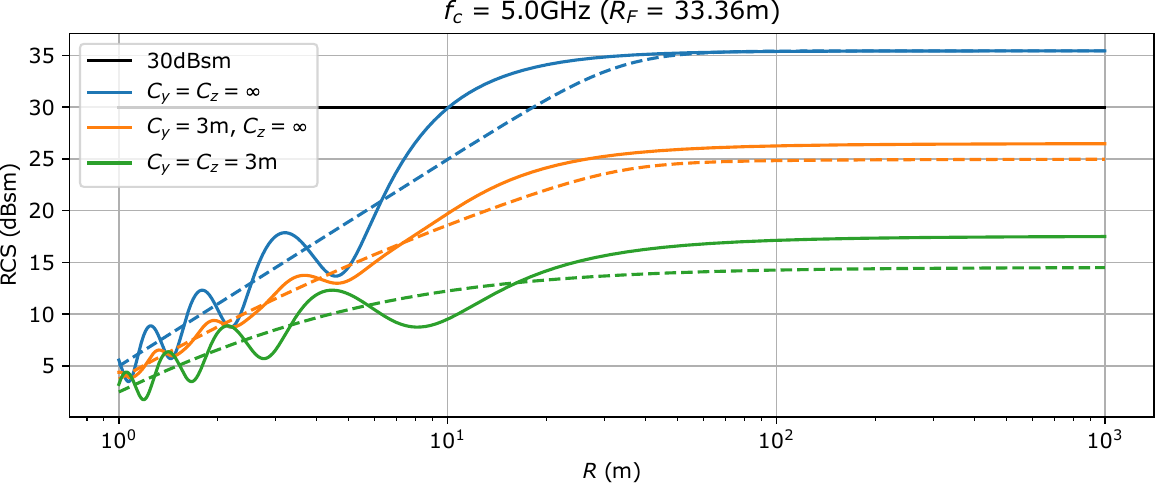}
\vspace{0.2cm}

\includegraphics[width=0.9\linewidth]{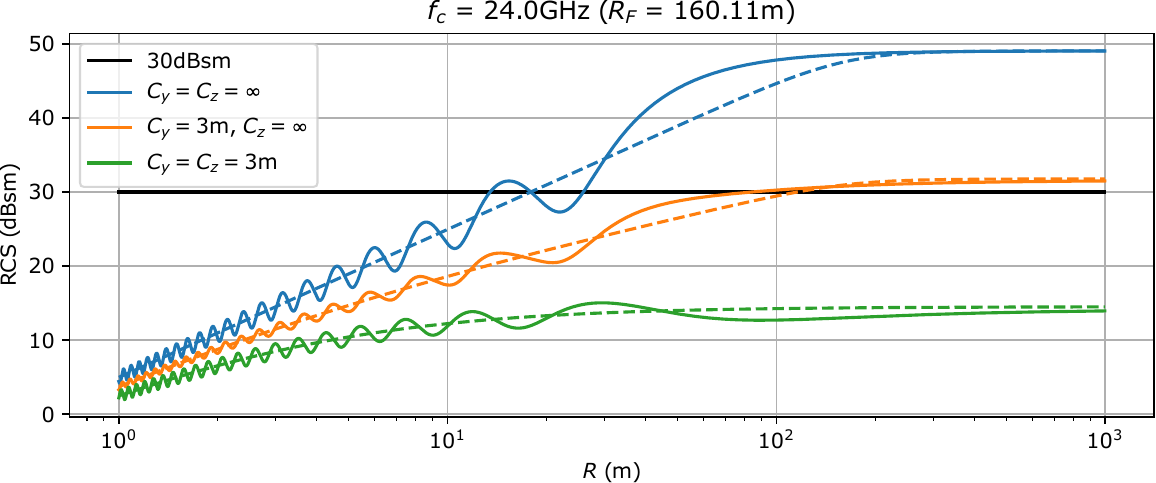}
\vspace{0.2cm}

\includegraphics[width=0.9\linewidth]{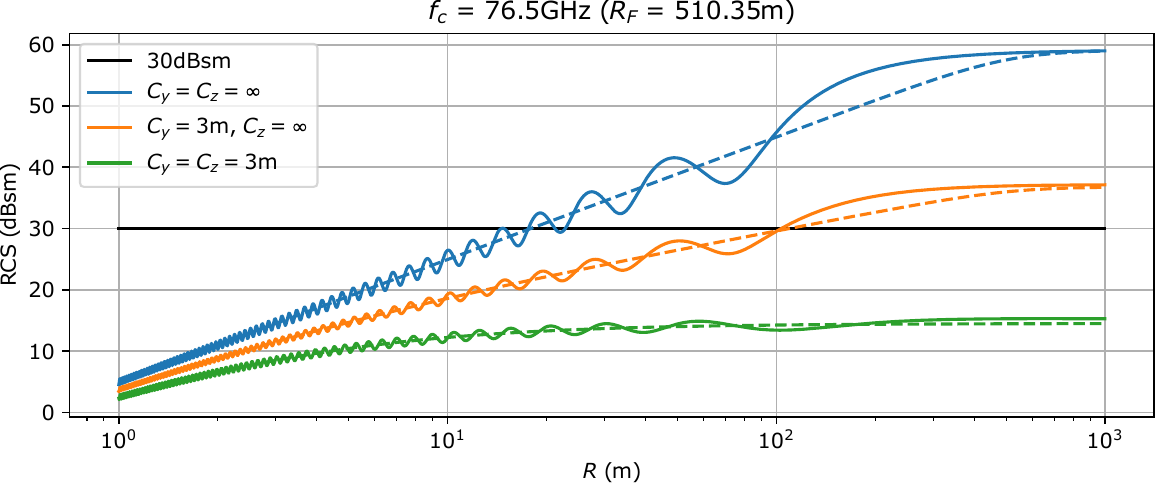}
\caption{RCS against distance for different carrier frequencies and different curvature radii. Infinite curvature radii correspond to a flat plate. The dashed lines represent the approximation obtained using \eqref{eq:RCS_curved_plate_approx} with $n=4$.}
\label{fig:RCS_curved_plate}
\end{figure}

Figure \ref{fig:RCS_curved_plate} shows the RCS against the distance for different carrier frequencies (5GHz for WiFi frequencies, 24GHz and 76.5GHz for automotive radars) and different curvature radii, with and without approximations. The higher the carrier frequency, the better the approximations. Compared to the flat plate geometry, introducing curvatures in the $y$ and $z$ axis modifies drastically the behaviour of the RCS. First, when curvature radii are introduced in both dimensions such that $C_y << R_\text{F}$ and $C_z << R_\text{F}$, the RCS is constant when $R >> C_y$ and $R >> C_z$, whatever the Fraunhoffer distance. In the case where $C_y << R << C_z$ (or $C_z << R << C_y$ by symmetry), the RCS is proportional to the distance, until the Fraunhoffer distance is reached. Finally, in the case where $R << C_y$, $R << C_z$ and $R << R_\text{F}$, the RCS is proportional to the square of the distance. At high carrier frequencies, the Fraunhoffer distance is high compared to the considered distances in indoor and automotive applications. Therefore, the modelling of the target has a huge impact on the radar propagation model, since the RCS is either constant, proportional to $R$ or proportional to $R^2$ depending on the geometry.

\section{EM-based radar propagation model}
\label{sec:radar_prop_models}

Based on the results obtained in Section \ref{sec:EM}, multiple models can be defined to evaluate the received power in free space: 
\begin{enumerate}
    \item The ray-tracing model, which computes the received power based on geometrical optics as 
    \begin{equation}
        P_{\R,\text{RT}} = G_r \frac{\lambda^2}{4\pi} \: \frac{\text{EIRP}}{4\pi} \: \frac{1}{(2R)^2}, \label{eq:PR_RT2}
    \end{equation}
    presented in Section \ref{sec:SOTA_prop} and obtained from \eqref{eq:SEF_RT}.
    \item The constant RCS model, which computes the received power in far field following the radar equation:
    \begin{equation}
        P_{\R,\text{RCS}} = G_r \frac{\lambda^2}{4\pi} \: \frac{\text{EIRP}}{(4\pi)^2} \: \frac{1}{R^4} \: \sigma, \label{eq:PR_RCS2}
    \end{equation}
    presented in Section \ref{sec:SOTA_prop}, and obtained from \eqref{eq:SEF_RCS} in far field, with $\sigma$ the RCS of the scatterer, independent of the distance $R$. 
    \item The new radar propagation model which computes the received power as 
    \begin{align}
        &P_\R = G_r \frac{\lambda^2}{4\pi} \: \frac{\text{EIRP}}{4\pi} \: \frac{1}{(2R)^2} \: \frac{\tilde{R}_y\tilde{R}_z}{R^2}\left|\Gamma\left(\frac{\tilde{R}_y}{R_\text{F}}\right)\Gamma\left(\frac{\tilde{R}_z}{R_\text{F}}\right)\right| \label{eq:PR_new_RT} \\
        &= G_r \frac{\lambda^2}{4\pi} \: \frac{\text{EIRP}}{(4\pi)^2} \: \frac{1}{R^4} \: \underbrace{\pi \tilde{R}_y\tilde{R}_z \left|\Gamma\left(\frac{\tilde{R}_y}{R_\text{F}}\right)\Gamma\left(\frac{\tilde{R}_z}{R_\text{F}}\right)\right|}_{\sigma}. \label{eq:PR_new_RCS} \\
        \nonumber &\approx G_r \frac{\lambda^2}{4\pi} \: \frac{\text{EIRP}}{4\pi} \: \frac{R_\F^2}{4R^4} \\
        & \quad\qquad\cdot\left[\bigg(1 + \bigg(\frac{R_\F}{\tilde{R}_y}\bigg)^n\bigg)\bigg(1+\bigg(\frac{R_\F}{\tilde{R}_z}\bigg)^n\bigg)\right]^{-\frac{1}{n}},
    \end{align}
    computed from \eqref{eq:SEF_new}, and using the approximation of \eqref{eq:gamma_approx}. Equation \eqref{eq:PR_new_RT} is equivalent to \eqref{eq:PR_RT2} with a correction factor, or \eqref{eq:PR_new_RCS} to \eqref{eq:PR_RCS2} in which the definition of the distance-dependent RCS given in \eqref{eq:RCS_curved_plate} has been included.
\end{enumerate}
These models can also be generalised to be compatible with any path-loss model with intercept $\beta$ and exponent $\alpha$. In that case, \eqref{eq:PR_RT2}, \eqref{eq:PR_RCS2}, \eqref{eq:PR_new_RT} and \eqref{eq:PR_new_RCS} respectively become
\begin{alignat}{10}
    &P_{\R,\text{RT}} = G_r \frac{\lambda^2}{4\pi} \: \frac{\text{EIRP}}{\beta} \: \frac{1}{(2R)^\alpha}, \label{eq:PR_RT_full} \\
    &P_{\R,\text{RCS}} = G_r \frac{\lambda^2}{4\pi} \: \frac{\text{EIRP}}{\beta^2} \: \frac{1}{R^{2\alpha}} \: \sigma, \label{eq:PR_RCS_full} \\
    &P_\R = G_r \frac{\lambda^2}{4\pi} \: \frac{\text{EIRP}}{\beta} \: \frac{1}{(2R)^\alpha} \: \frac{(\tilde{R}_y\tilde{R}_z)^{\frac{\alpha}{2}}}{R^\alpha}\left|\Gamma\left(\frac{\tilde{R}_y}{R_\text{F}}\right)\Gamma\left(\frac{\tilde{R}_z}{R_\text{F}}\right)\right|^{\frac{\alpha}{2}} \\
    &= G_r \frac{\lambda^2}{4\pi} \: \frac{\text{EIRP}}{\beta^2} \: \frac{1}{R^{2\alpha}} \: \underbrace{\frac{\beta(\tilde{R}_y\tilde{R}_z)^{\frac{\alpha}{2}}}{2^\alpha}\left|\Gamma\left(\frac{\tilde{R}_y}{R_\text{F}}\right)\Gamma\left(\frac{\tilde{R}_z}{R_\text{F}}\right)\right|^{\frac{\alpha}{2}}}_{\sigma} \label{eq:PR_RCS_new_full} \\
    \nonumber &\approx G_r \frac{\lambda^2}{4\pi} \: \frac{\text{EIRP}}{\beta} \: \left(\frac{R_\F}{2R^2}\right)^\alpha \\
    &\qquad\qquad \cdot \left[\bigg(1 + \bigg(\frac{R_\F}{\tilde{R}_y}\bigg)^n\bigg)\bigg(1+\bigg(\frac{R_\F}{\tilde{R}_z}\bigg)^n\bigg)\right]^{-\frac{\alpha}{2n}}. \label{eq:PR_RCS_new_full_approx}
\end{alignat}
Note that the RCS definition in \eqref{eq:RCS_curved_plate} must be generalised to take into account the path-loss-model:
\begin{align}
\sigma &= \beta R^\alpha \frac{|\vec{E}_s|^\alpha}{|\vec{E}_i|^\alpha} = \frac{\beta(\tilde{R}_y\tilde{R}_z)^{\frac{\alpha}{2}}}{2^\alpha}\left|\Gamma\left(\frac{\tilde{R}_y}{R_\text{F}}\right)\Gamma\left(\frac{\tilde{R}_z}{R_\text{F}}\right)\right|^{\frac{\alpha}{2}} \\
&\quad\approx \frac{\beta R_\F^\alpha}{2^\alpha} \left[\bigg(1 + \bigg(\frac{R_\F}{\tilde{R}_y}\bigg)^n\bigg)\bigg(1+\bigg(\frac{R_\F}{\tilde{R}_z}\bigg)^n\bigg)\right]^{-\frac{\alpha}{2n}}.
\end{align}

%% file: numerical_analysis.tex
\begin{table}
    \centering
    \caption{Scenario's parameters.}
    \label{tab:scenarios_parameters}
    \renewcommand{\arraystretch}{1.2}
    \begin{tabular}{|lc|lc|}
    \hline
    Vehicles' density & 10$\text{km}^{-1}$ & Plate side length $a$ & 1m  \\\hline
    Number of lanes & 5 & Lanes' width & 3.6m \\\hline
    Interference probability & 0.01 & Path-loss exponent $\alpha$ & 2 \\\hline
    Path-loss intercept $\beta$ & $4 \pi$ & Carrier frequency & 76.5GHz \\\hline 
    EIRP & 10dB & Receiver gain $G_r$ & 30dBi\\\hline 
    Antenna beamwidth & 15\degree & SINR threshold  & 10dB \\\hline
    \end{tabular}
    \renewcommand{\arraystretch}{1}
\end{table}

In this section, the different radar propagation models developed in Section \ref{sec:radar_prop_models} are compared within the stochastic geometry framework, following the scenario studied in \cite{9119440}. In that paper, the authors study the performance of automotive radar systems in a bidirectional multi-lane road. Instead of simply considering a constant RCS, they model fluctuation by introducing a random variable following a Chi-square distribution, as suggested in the Swerling I and III models. The performance metric is the radar success probability, defined as the probability that the radar SINR is higher than a given SINR threshold for at a given distance. In this work, the considered parameters for the scenario are summarised in Table \ref{tab:scenarios_parameters}. Random fluctuation models can be coupled with the proposed radar propagation models by putting the models of Section \ref{sec:radar_prop_models} as the average RCS of the random variable distribution.\smallskip

\begin{figure}
\centering
\includegraphics[width=0.9\linewidth]{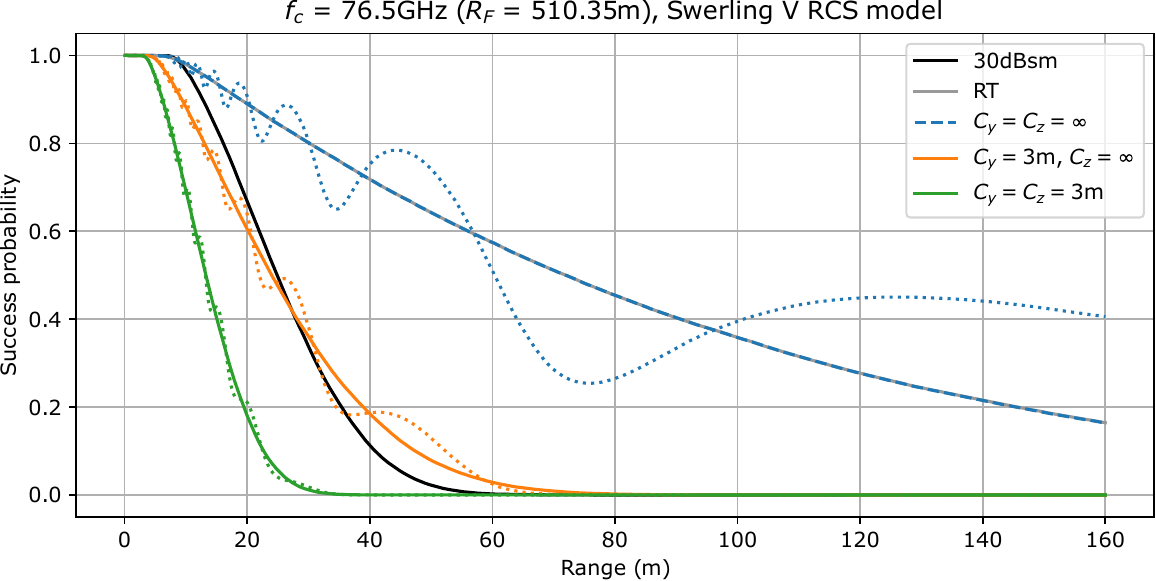}
\vspace*{0.2cm}

\includegraphics[width=0.9\linewidth]{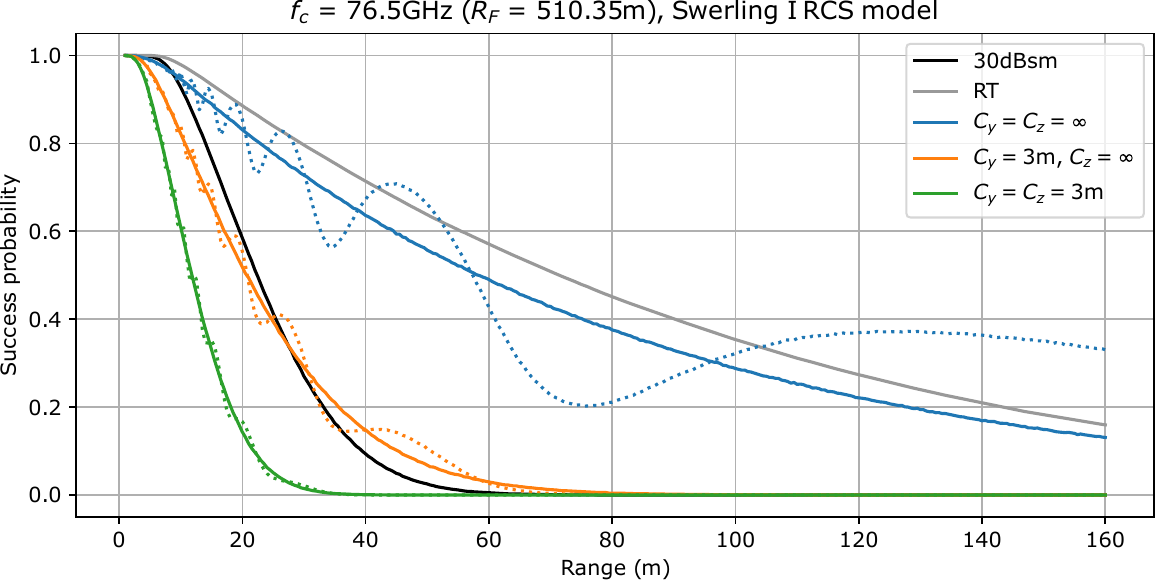}
\caption{Success probability achieved with the multiple radar propagation models in \eqref{eq:PR_RT_full} (gray), \eqref{eq:PR_RCS_full} (black), and \eqref{eq:PR_RCS_new_full} with different curvature radii (other colors, dotted lines). The approximations obtained with \eqref{eq:PR_RCS_new_full_approx} are also shown (other colors, solid lines). Swerling I (exponential distribution) and Swerling V (constant value) RCS fluctuation models are respectively considered in the bottom and top figures.}
\label{fig:success_probability}
\end{figure}

Figure \ref{fig:success_probability} illustrates the success probability achieved with the multiple radar propagation models of Section \ref{sec:radar_prop_models}, with the Swerling I (exponential distribution) and Swerling V (constant value) RCS fluctuations models. One can observe that the selected radar propagation model hugely impacts the radar performance since the received power varies a lot depending on the model. Additionally, among the new models, the choice of the curvature radii has also a huge impact on the achieved performance. While a flat square plate would be more appropriate for trucks, the introduction of a curvature radius in the vertical dimension seems more suited to cars, and leads to a large performance degradation. The ray-tracing model is close to the flat square model since a high carrier frequency is considered, leading to a high Fraunhoffer distance compared to the radar-to-target distance. Additionally, for the selected configuration, the model with a constant RCS of 30dBsm also approximates well the model with vertical curvature radius. In summary, the model should really be chosen differently depending on the targets' geometry.

%% file: conclusion.tex
In this paper, new radar propagation models and RCS modellings based on the EM theory have been developed. The targets are modelled as flat and curved square plates. For the flat square plate geometry, it helps to make the link between the geometrical optics propagation models used in ray-tracing applications, and the usual radar equation. Finally, these models have been applied to an automotive scenario initially studied within the stochastic geometry framework to evaluate the discrepancies between the different models. It has been shown that the model should be chosen accurately w.r.t. the radar target to well evaluate the performance of the radar system. In future works, other geometries will be studied as the flat disk and the sphere. Additionally, materials which are not PEC will be considered. Finally, measurements will be performed to validate the models. 